\documentstyle[prb,aps,floats,epsf]{revtex}

\draft

\begin{document}

\twocolumn[\hsize\textwidth\columnwidth\hsize\csname @twocolumnfalse\endcsname

\title{Plaquette-singlet solid state and topological hidden order \\ in
spin-1 antiferromagnetic Heisenberg ladder}

\author{Synge Todo$^{1,2,*}$, Munehisa Matsumoto,$^{1,\dagger}$ Chitoshi
Yasuda,$^{1,**}$ and Hajime Takayama$^{1,\dagger\dagger}$}

\address{$^1$Institute for Solid State Physics, University of Tokyo, Kashiwa
  227-8581, Japan}
\address{$^2$Theoretische Physik, Eidgen\"ossische Technische Hochschule,
   CH-8093 Z\"urich, Switzerland}

\date{\today}
\maketitle

\begin{abstract}
 Ground-state properties of the spin-1 two-leg antiferromagnetic ladder
 are investigated precisely by means of the quantum Monte Carlo method.
 It is found that the correlation length along the chains and the spin
 gap both remain finite regardless of the strength of interchain
 coupling, i.e., the Haldane state and the spin-1 dimer state are
 connected smoothly without any quantum phase transitions between them.
 We propose a plaquette-singlet solid state, which qualitatively
 describes the ground state of the spin-1 ladder quite well, and also a
 corresponding topological hidden order parameter.  It is shown
 numerically that the new hidden order parameter remains finite up to
 the dimer limit, though the conventional string order defined on each
 chain vanishes immediately when infinitesimal interchain coupling is
 introduced.
\end{abstract}


\vspace*{1.0em}

]\narrowtext

\section{Introduction}
\label{sec:intro}

Quantum spin-ladder systems have been studied theoretically and
experimentally over the last decade as materials with a novel {\em
spin-gap} state, as well as by their relevance to the high-temperature
superconductivity.~\cite{DagottoR1996} Especially, the two-leg ladder
Heisenberg antiferromagnet, which is defined by the Hamiltonian:
\begin{eqnarray}
 \label{eqn:Hamiltonian}
 {\cal H}
 & = & 
  J \sum_{i} \left\{ {\bf S}_{1,i} \cdot {\bf S}_{1,i+1} 
  + {\bf S}_{2,i} \cdot {\bf S}_{2,i+1} \right\} \nonumber \\
 & & 
  + K \sum_{i} {\bf S}_{1,i} \cdot {\bf S}_{2,i} ,
\end{eqnarray}
has been studied most extensively.  Here, ${\bf S}_{\alpha,i}$ is the
spin operator at site $i$ on the $\alpha$-th chain ($\alpha=1,2$), and
the intrachain and interchain coupling constants are denoted by $J$ and
$K$, respectively.  In the following, we restrict our attention only to
the case in which the intrachain coupling is antiferromagnetic ($J>0$).
On the other hand, the interchain coupling constant, $K$, can be either
positive (antiferromagnetic) or negative (ferromagnetic).

At $K=0$, the system consists of two decoupled antiferromagnetic
Heisenberg chains.  In this case, it is well known that the ground-state
properties can be classified into two universality classes depending on
the parity of $2S$.  Here $S$ is the spin size.  In the case where $S$
is a half-odd integer, the ground state is {\em critical}, i.e., the
system has gapless low-lying excitation and the antiferromagnetic
correlation function along the chain decays in an algebraic way as the
distance increases.  On the other hand, it is conjectured by
Haldane\cite{Haldane1983} that the antiferromagnetic Heisenberg chain of
integer spins has a finite excitation gap above its {\em unique} ground
state, and the correlation function decays exponentially with a finite
correlation length.  Its ground-state properties can be understood quite
well from the viewpoint of the valence-bond solid (VBS)
picture,\cite{AffleckKLT1987} in which the ground state is essentially
represented as direct products of spin-$\frac{1}{2}$ dimers (AKLT
state).  In addition, a topological order parameter characterizing the
AKLT state, as well as the Haldane state, so-called the {\em string
order parameter}, has been proposed.\cite{denNijsR1989} The validity of
Haldane's conjecture has been confirmed precisely for $S=1$, 2, and 3 by
several numerical
methods.\cite{NightingaleB1986,WhiteH1993,GolinelliJL1994,%
SchollwockJ1995,TodoK2001}

Introduction of non-zero interchain coupling, $K$, is known to
drastically change the ground state, at least for the spin-$\frac{1}{2}$
case.\cite{DagottoR1996} For small $K$, either antiferromagnetic or
ferromagnetic, it immediately opens a spin gap of $O(|K|)$ with some
logarithmic corrections.~\cite{TotsukaS1995} That is, $K=0$ is the
special point at which there occurs a quantum second-order phase
transition between the dimer phase ($K>0$) and the spin-1 Haldane phase
($K<0$).  Again, from the viewpoint of VBS picture, one can understand
this phase transition as a global rearrangement of dimer pattern.  For
larger half-odd-integer spins ($S=\frac{3}{2}$, $\frac{5}{2}$,
$\cdots$), the criticality at $K=0$ should be essentially the same as in
the spin-$\frac{1}{2}$ case.

In the case of {\em integer-spin} chains, on the other hand, effects of
interchain coupling have been known little so far.  Recently,
S\'en\'echal and Allen studied the spin-1 ladder by mapping it to the
nonlinear $\sigma$ model\cite{Senechal1995} and also by the bosonization
technique.\cite{AllenS2000} They found that in contrast to the
spin-$\frac{1}{2}$ case, small interchain coupling reduces the magnitude
of the spin gap in both of antiferromagnetic and ferromagnetic regimes.
In addition, their analyses as well as their complemental Monte Carlo
calculation suggest that there is no critical point between the Haldane
and the spin-1 dimer phases.  This may seem paradoxical since these two
phases have apparently different dimer patterns from each other.

In this paper, we present the results of our extensive quantum Monte
Carlo simulation on the spin-1 ladder.  After reviewing details of our
simulation using the efficient continuous-time loop algorithm in
Sec.~\ref{sec:method}, we present our numerical data on the uniform
susceptibility, staggered susceptibility, antiferromagnetic correlation
length, etc. in Sec.~\ref{sec:results}, which convincingly demonstrate
the continuity of the two limiting case ($K=0$ and $K=\infty$), and thus
support the conjecture by the previous analytical
approaches.\cite{Senechal1995,AllenS2000} In Sec.~\ref{sec:pss}, we
propose a {\it plaquette-singlet solid state}, which is constructed as
products of local singlet states of four $S=\frac{1}{2}$ spins.  The
Haldane state and the spin-1 dimer state are naturally included as
special limits.  In addition, we propose a kind of hidden order
parameter, which can detect the topological hidden order exsisting in
the plaquette-singlet solid state.  We show numerically that the new
hidden order parameter we propose remains finite in the whole parameter
range, $0 \le K \le \infty$, while the conventional string order
parameter vanishes except at $K=0$.  In Sec.~\ref{sec:ferro}, we
consider in turn the case where the interchain coupling is ferromagnetic
($K<0$), and show that the spin-1 Haldane state and that of $S=2$ also
continue to each other without any singularity on the way to the other.
We give a summary of our results and some discussions in the final
section.

\section{Quantum Monte Carlo Method}
\label{sec:method}

\subsection{Continuous-time loop algorithm for spin-1 system}
\label{subsec:loop}

The recently-developed continuous-time loop
algorithm~\cite{EvertzLM1993,WieseY1994,BeardW1996} is one of the most
efficient methods for simulating quantum spin systems. It is a variant
of the world-line Monte Carlo method, which is based on the
path-integral representation by means of the Suzuki-Trotter
discretization.~\cite{Suzuki1994} However, the continuous-time loop
algorithm works directly in the imaginary-time
continuum,~\cite{BeardW1996} and thus is completely free from the
systematic error in the Suzuki-Trotter discretization. In addition, the
correlation between successive spin configurations is greatly reduced,
sometimes by orders of magnutide, since it flips effectively clusters of
spins, or {\em loops}, whose linear sizes correspond directly to the
length scale of relevant spin fluctuations.  The algorithm has already
been applied to various spin systems with great
success.~\cite{Evertz1997}

\begin{figure}[tb]
 \centerline{\epsfxsize=0.32\textwidth\epsfbox{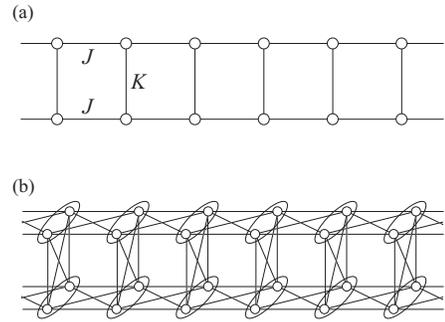}}
 \vspace*{1.5em}
 \caption{(a) Original spin-1 ladder, and (b) equivalent system
 represented by $S=\frac{1}{2}$ spins (subspins).  The spin-1 ladder of
 length $L$ ($2L$ spins and $3L$ bonds) is mapped onto the
 spin-$\frac{1}{2}$ system on a lattice of $4L$ subspins and $12L$
 bonds.  Each oval in (b) denotes a pair of subspins which is
 symmetrized by being applied special boundary conditions in the
 imaginary-time direction.}
 \label{fig:lattice}
\end{figure}

The Hamiltonian we consider is given by Eq.~(\ref{eqn:Hamiltonian}) with
$S=1$.  The linear size along the chain is denoted by $L$, and we adopt
periodic boundary conditions in this direction, i.e., ${\bf
S}_{\alpha,i+L} = {\bf S}_{\alpha,i}$ for $\alpha = 1$ and 2.  In order
to apply the continuous-time loop algorithm to the present spin-1
system, first we represent the spin-1 Hamiltonian in terms of subspins.
In this representation, each spin-1 operator in
Eq.~(\ref{eqn:Hamiltonian}) is decomposed into a sum of two
spin-$\frac{1}{2}$ operators.~\cite{KawashimaG1994} Simultaneously, each
bond of strength $J$ (or $K$) is transformed into four bonds of the same
strength connecting subspins.  The lattices before and after the subspin
transformation are shown in Fig.~\ref{fig:lattice}.  Note that in order
to recover dimensions of the original spin-1 Hilbert space ($3^{2L}$),
one needs to introduce a set of projection operators, each of which acts
on a pair of subspins and projects out the state with $S=0$
(Fig.~\ref{fig:lattice}).  After transformed into a path-integral
representation, the projection operators are converted to special
boundary conditions in the imaginary-time
direction;~\cite{TodoK2001,HaradaTK1998} for each pair of subspins, the
total $S^z$ is required to be conserved across the imaginary-time
boundary.

For the mapped system, the spin-$\frac{1}{2}$ continuous-time loop
algorithm~\cite{EvertzLM1993,WieseY1994,BeardW1996} can be applied
without any modification except that we need to introduce additional
graphs and labeling rules for the boundaries in the imaginary-time
direction.~\cite{TodoK2001,HaradaTK1998} We use the multi-cluster
variant of the loop algorithm.  The resulting algorithm is found to work
quite well as the same as the original algorithm developed for
$S=\frac{1}{2}$; the integrated auto-correlation time for the physical
quantities we measure remains of order unity, and there is observed no
significant sign of its growth even in the largest system in the present
simulation ($L=1024$ and $T/(J+K) = 2/1024 \simeq 0.00195$).

\subsection{Physical quantities}
\label{subsec:quantity}

The physical quantities of interest can be measured by using the
corresponding subspin representations.  First, for later convenience, we
introduce the {\it imaginary-time} dynamical correlation function,
$C^{\pm}(x,\tau)$, and its Fourier transform, ${\tilde
C}^{\pm}(k,\omega)$.  The former is defined explicitly in the
path-integral representation by
\begin{eqnarray}
 \label{eqn:dstr}
 C^{\pm}(x,\tau) & = & \frac{1}{2 L \beta} \Big< \int_0^\beta \!\!\!\!
 dt \sum_{i} 
  \left\{ {\cal S}_{1,i}(t) \pm {\cal S}_{2,i}(t) \right\} \nonumber \\
 & & \quad \times \left\{ {\cal S}_{1,i+x}(t+\tau) \pm {\cal S}_{2,i+x}(t+\tau) \right\} \Big> \,,
\end{eqnarray}
where $\beta$ is the inverse temperature ($=1/T$).  The spin
configuration at site $i$ on the $\alpha$-th chain and at imaginary time
$\tau$ $(0 \le \tau \le \beta)$ is denoted by ${\cal
S}_{\alpha,i}(\tau)$, which takes -1, 0, or 1.  The bracket $\left<
\cdots \right>$ in Eq.~(\ref{eqn:dstr}) denote the average over Monte
Carlo steps (MCS).  In the present subspin representation, ${\cal
S}_{\alpha,i}(\tau)$ is simply given by a sum of $S^z$ of two subspins
at $(\alpha,i,\tau)$.

In terms of the imaginary-time dynamic structure factor,
$\tilde{C}^\pm(k,\omega)$, the uniform susceptibility and the staggered
susceptibility are simply given by
\begin{equation}
 \chi = 2L\beta \tilde{C}^{+}(0,0)
\end{equation}
and
\begin{equation}
 \chi_{\rm s} = 2L\beta \tilde{C}^{-}(\pi,0),
\end{equation}
respectively.  We also calculate the static structure factor at
momentum~$\pi$ as
\begin{eqnarray}
 \label{eqn:sstr}
 S(\pi) &=& \frac{1}{2L} \Big< \sum_{i,j} (-1)^{|i-j|} \big( {\cal S}_{1,i}(0) - {\cal S}_{2,i}(0) \big) \nonumber \\ 
 & & \quad \quad \times \big( {\cal S}_{1,j}(0) - {\cal S}_{2,j}(0) \big) \Big>.
\end{eqnarray}
In order to calculate the correlation length along the chains, we use
the second-moment method:~\cite{CooperFP1982}
\begin{equation}
 \label{eqn:corr}
 \xi = \frac{L}{2\pi}
  \sqrt{\frac{\tilde{C}^{\pm} (\pi,0)}{\tilde{C}^{\pm} (\pi+2\pi/L,0)}-1} .
\end{equation}
Similarly, the spin gap, which is defined as the inverse of the
correlation length in the imaginary-time direction, is measured by
\begin{equation}
 \label{eqn:gap}
 \Delta^{-1} = \frac{\beta}{2\pi}
  \sqrt{\frac{\tilde{C}^{\pm}(\pi,0)}{\tilde{C}^{\pm}(\pi,2\pi/\beta)}-1} .
\end{equation}
In Eqs.~(\ref{eqn:corr}) and (\ref{eqn:gap}), we take the minus (plus)
sign for $K>0$ ($K<0$).  Although the above second-moment estimates
suffer from systematic error due to the existence of subdominant
decaying modes in the correlation function, it should be sufficiently
small (at $K=0$ the systematic error for the spin gap is known to be
about 0.2\%~\cite{TodoK2001}), and thus we expect that it would be
irrelevant to the following discussions.  We will also present our
results for the string order parameter~\cite{denNijsR1989} and a new
hidden order parameter in Sec.~\ref{sec:pss}.  Their explicit
definitions will be given later.

\begin{table}[tb]
 \caption{Convergence of physical quantities at $J=0.7$ and $K=0.3$.
 The temperature is taken as $T/(J+K)=2/L$ for each $L$.  The figure in
 parentheses denotes the statistical error ($2\sigma$) in the last
 digit.  There are observed no significant differences between the data
 with $L=256$, 512, and 1024.}
 \label{tab:convergence}
 \begin{tabular}{rrrrrr}
  \multicolumn{1}{c}{$L$} & 
  \multicolumn{1}{c}{MCS} & 
  \multicolumn{1}{c}{$(J+K)\,\chi_{\rm s}$} &
  \multicolumn{1}{c}{$S(\pi)$} &
  \multicolumn{1}{c}{$\Delta/(J+K)$} &
  \multicolumn{1}{c}{$\xi$}
  \\
  \hline
  8 & $3\times10^5$ & 16.48(3)     &  4.95(1) &  0.3863(6)\ \, &  4.57(1) \\
  16 & $6\times10^5$ & 49.72(8)     &  7.74(1) &  0.2120(3)\ \, &  8.50(1) \\
  32 & $7\times10^5$ & 138.0(2)\ \, & 11.39(2) &  0.1218(2)\ \, & 14.86(2) \\
  64 & $1\times10^6$ & 315.9(4)\ \, & 14.97(2) &  0.0793(1)\ \, & 22.81(3) \\
  128 & $1\times10^6$ & 468.7(8)\ \, & 16.23(2) &  0.06602(9)    & 27.37(4) \\
  256 & $7\times10^5$ & 489.8(5)\ \, & 16.19(2) &  0.06487(7)    & 27.91(4) \\
  512 & $3\times10^5$ & 490.3(3)\ \, & 16.17(2) &  0.06477(8)    & 27.93(3) \\
  1024& $2\times10^5$ & 490.1(2)\ \, & 16.17(2) &  0.06476(4)    & 27.92(2) \\
 \end{tabular}
\end{table}

In practice, all the physical quantities we will show in the following,
including the hidden order parameters, are measured by using so-called
{\em improved estimators}.  For example, the staggered susceptibility
for $K>0$ is simply represented as the sum of squared length of each
loop, divided by $8 L \beta$.  Similarly, the imaginary-time dynamic
structure factor can be measured directly as
\begin{eqnarray}
 \label{eqn:imp}
{\tilde C}^{\pm}(k,\omega) = \frac{1}{4 L^2 \beta^2} \left< \sum_p \left| 
\frac{1}{2} \oint (\pm 1)^\alpha e^{i(k x+\omega t)} d \ell \right|^2 \right>,
\end{eqnarray}
where the integration is performed along a closed path on each loop, and
the summation runs over all loops.

\subsection{Taking the thermodynamic limit and the zero-temperature limit}
\label{subsec:limit}

In the present method, the system size, $L$, and the temperature, $T$,
are restricted to be finite.~\cite{EvertzL2000} Since we are mainly
interested in the ground-state properties of the infinite lattice, a
proper extrapolation scheme for taking the thermodynamic limit ($L
\rightarrow \infty$) as well as the zero-temperature one ($T \rightarrow
0$) is required.  In the present study, we adopt the following strategy,
which is the same as was used in Ref.~\onlinecite{TodoK2001} for the
estimation of the Haldane gap of the antiferromagnetic Heisenberg
chain with $S=1$, 2, and 3.

For each parameter set $(J,K)$, we start with a small lattice at
relatively high temperature, e.g., $L=8$ and $T/(J+K)=0.25$.  Then, we
increase the system size exponentially step by step, like $L=16$, 32,
64, 128, $\cdots$.  Simultaneously, the temperature is decreased so as
to keep $LT/(J+K) = L/(J+K)\beta$ constant.  In other words, the {\em
aspect ratio} of the (1+1)-dimensional space is kept unchanged for all
$L$'s.  If the system is gapless, the {\em finite-size scaling} holds;
the correlation length in the real-space direction, $\xi(L,T)$, and that
in the imaginary-time direction, $\Delta^{-1}(L,T)$, both would grow
being proportional to $L$.~\cite{z} Simultaneously, the other physical
quantities, such as the staggered susceptibility, should exhibit
power-law behavior with some exponents depending on their own anomalous
dimensions.

On the other hand, if the system is {\em gapful} (this is the case for
the present system as we will see below), there exist finite intrinsic
correlation lengths, $\xi$ and $\Delta^{-1}$.  As long as the system
size and the inverse temperature are smaller enough than these intrinsic
correlation lengths, the critical behavior above mentioned is still
observed.  However, once both of $L$ and $\beta$ exceed them enough,
$\xi(L,T)$ and $\Delta(L,T)$, as well as other physical quantities, no
longer exhibit system-size dependence.  Strictly speaking, the
systematic error due to the finiteness of the system decreases
exponentially as the system size increases, and it becomes much smaller
than the statistical error due to the finiteness of MCS.

In Table~\ref{tab:convergence}, we show the system-size (and
temperature) dependence of the physical quantities at $(J,K) =
(0.7,0.3)$.  As seen clearly, the data with $L \ge 256$ (and with
$T/(J+K) \le 2/256 = 0.0078125$) exhibit no system-size dependence.
Thus, in this case, one can safely conclude that the system is gapful
and also that the physical quantities obtained are those of the infinite
lattice at zero temperature besides the statistical error.  Empirically,
we find that $L/\xi(L,T) > 6$ with $\beta \Delta(L,T) > 6$ is a
reasonable condition to guarantee the convergence in the present
numerical accuracy.

One of the advantages of the present scheme is that it depends on {\em
no} numerical extrapolation techniques, such as least-squares fitting,
the Shanks transform, etc.  Final results are simply obtained from those
of the largest system at the lowest temperature in the simulation.
Therefore, this method is quite stable and the error estimation is also
quite reliable.  It should be emphasized that the final results are not
affected at all by the value chosen for the aspect ratio, $L/(J+K)\beta$
(=2 in the present case).  However, if one chooses a too small or too
large value, the physical memory of the computer system might be
exhausted before reaching the thermodynamic limit or the
zero-temperature one.

In what follows, we will mainly present the data with $L=256$ and
$T/(J+K)=2/256$ unless otherwise noted.  Measurement of physical
quantities is performed for $5 \times 10^5 \sim 1 \times 10^6$ MCS after
$10^3$ MCS for thermalization.  Typically, simulation of this system
requires about 7MB of physical memory, and 1 MCS takes about
0.33~sec. on a single CPU of SGI 2800 (MIPS R12000 400MHz).  This system
size is somewhat exaggerated for certain sets of parameters, $(J,K)$.
However, in such cases, the system can be considered as ${\cal N}$ {\em
statistically-independent} samples simulated in parallel, where ${\cal
N} \sim (L/\xi) \times (\Delta \beta)$.  Therefore, we gain better
statistics proportional to $\sqrt{\cal N}$, which completely compensates
for the growth of CPU time ($\sim L \beta$ for large $L$ and
$\beta$).~\cite{statistics} Thus, we loose nothing besides the memory
requirement.  This is already manifested in the figures presented in
Table~\ref{tab:convergence}.

\section{Numerical Results}
\label{sec:results}

\subsection{Parameterization}
\label{subsec:param}

The ground state of the present system is parametrized by the ratio of
the interchain coupling constant to the intrachain one, $x \equiv K/J$.
Hereafter, we mainly use $R$, which is defined by
\begin{equation}
R = \frac{K}{J+|K|} = \frac{x}{1+|x|} \,.
\end{equation}
Since we consider only the antiferromagnetic intrachain coupling ($J \ge
0$), $-1 \le R \le 1$.  At $R=0$, the system consists of two independent
antiferromagnetic chains (spin-1 Haldane chains).  On the other hand, at
$R=1$, it is decoupled into dimers sitting on each rung.  The limit $R
\rightarrow -1$ corresponds to a single spin-2 antiferromagnetic chain.
We also introduce the reduced temperature, $\tilde{T} = T/(J+K)$,
reduced susceptibilities, $\tilde{\chi} = (J+K) \chi$ and
$\tilde{\chi_{\rm s}} = (J+K) \chi_{\rm s}$, and the reduced gap,
$\tilde{\Delta} = \Delta/(J+K)$.  In this section and
Sec.~\ref{sec:pss}, we consider only the case where $0 \le R \le 1$,
i.e., the interchain coupling is antiferromagnetic.

\subsection{Uniform susceptibility}
\label{subsec:usus}

\begin{figure}[tb]
 \centerline{\epsfxsize=0.42\textwidth\epsfbox{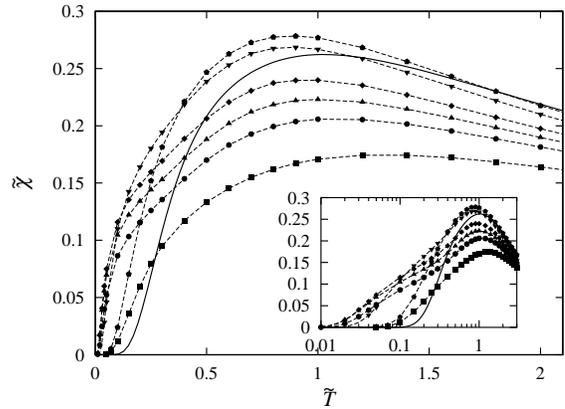}}
 \vspace*{1.0em}
 \caption{Temperature dependence of the uniform susceptibility for $R=0$
 (squares), 0.2 (circles), 0.3 (upward triangles), 0.4 (diamonds), 0.6
 (downward triangles), and 0.8 (pentagons).  The exact result for $R=1$
 (Eq.~(\protect\ref{eqn:dimer})) is plotted by a solid line.  In the
 inset, the same data are plotted against the logarithm of $\tilde{T}$.
 }
 \label{fig:usus}
\end{figure}

Before investigating the zero-temperature properties of the spin-1
ladder, first we briefly discuss finite-temperature behavior of the
uniform susceptibility, which gives a rough profile on the excitation
spectrum of the system.  In Fig.~\ref{fig:usus}, $\tilde{\chi}$ is
plotted as a function of $\tilde{T}$ for $R=0$, 0.2, 0.3, 0.4, 0.6, 0.8,
and 1.  At $R=1$, i.e., $J=0$, since the system consists of independent
dimers, the exact form of the uniform susceptibility is easily obtained
as
\begin{eqnarray}
 \label{eqn:dimer}
  \tilde{\chi}
  &=& \frac{1}{\tilde{T}} \, \frac{\exp (-1/\tilde{T}) 
  + 5 \exp (-3/\tilde{T})}{1 + 3 \exp(-1/\tilde{T}) + 5 \exp(-3/\tilde{T})} 
 \nonumber \\
  & \simeq & \frac{1}{\tilde{T}} \, \exp (-1/\tilde{T}) \,\,\,\,\,\, \mbox{for $\tilde{T} \ll 1$.}
\end{eqnarray}
At $R=0$, $\tilde{\chi}$ is also known to vanish at low temperatures
exponentially as $\exp(- \tilde{\Delta} / \tilde{T})$ with
$\tilde{\Delta} \simeq 0.41050(2)$~\cite{WhiteH1993} besides a prefactor
of some powers of $T$.  As seen clearly in Fig.~\ref{fig:usus}, the
temperature dependence of the uniform susceptibility does not depend on
the value of $R$ strongly; it has a very broad peak around $\tilde{T}
\simeq 1$, and decreases quite rapidly at lower temperatures.  It
suggests {\em spin-singlet ground state} regardless of the value of $R$.

However, it should be noted that at temperatures lower than $\tilde{T}
\simeq 0.4$, the uniform susceptibility is {\em not} a monotonic
function of $R$.  It is greatly enhanced by orders of magnitude around
$R = 0.3$, in comparison with those at $R=0$ and 1.  It indicates that
in the intermediate region of $R$, the spin gap is strongly suppressed,
and on the other hand long-range antiferromagnetic fluctuations are
enhanced, due to the competition between the intrachain and the
interchain antiferromagnetic couplings.

In addition, in the temperature profile of the uniform susceptibility
for $0.2 \le R \le 0.6$, one can see a clear {\em shoulder} structure at
$\tilde{T} \simeq 0.1$ (see also the inset of Fig.~\ref{fig:usus}),
which indicates existence of some additional anomalies in the excitation
spectrum of the system.

\begin{figure}[tb]
 \centerline{\epsfxsize=0.42\textwidth\epsfbox{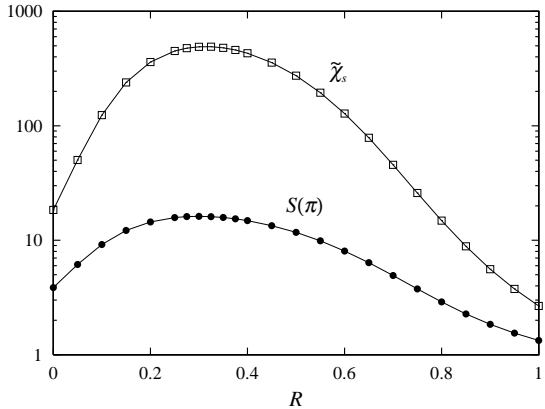}}
 \vspace*{1.0em}
 \caption{$R$-dependence of the staggered susceptibility (open squares)
 and the static structure factor at momentum $\pi$ (solid circles).  The
 statistical error of each data point is much smaller than the symbol
 sizes.}
 \label{fig:ssus}
\end{figure}

\subsection{Quantities at zero temperature}
\label{subsec:zero}

The temperature dependences of the staggered susceptibility,
$\tilde{\chi}_{\rm s}$, and the static structure factor, $S(\pi)$, are
found to be qualitatively very similar to those of the single Haldane
chain ($R=0$); they grow rapidly around $\tilde{T} \simeq 1$ and are
saturated to finite values at low temperatures, though the saturation
temperature strongly depends on $R$ (see the $R$-dependence of $\Delta$
shown below).  As for $S(\pi)$, in addition, it makes a weak peak before
saturated to the zero-temperature value, which indicates a short-range
antiferromagnetic order.

In Fig.~\ref{fig:ssus}, the zero-temperature values of
$\tilde{\chi}_{\rm s}$ and $S(\pi)$ are plotted as a function of $R$.
The value of $\tilde{\chi}_{\rm s}$ at $R=0$ is consistent with that
obtained in the previous work, $\tilde{\chi}_{\rm s} =
18.4048(7)$~\cite{TodoK2001}.  On the other hand, at $R=1$, they
coincide the exact values, $\tilde{\chi}_{\rm s} = \frac{8}{3}$ and
$S(\pi) = \frac{4}{3}$, respectively, within the statistical error.  One
sees in Fig.~\ref{fig:ssus} that they are smooth functions of $R$, and
thus there is no indication of singularities in the whole range of $R$.

\begin{figure}[tb]
 \centerline{\epsfxsize=0.42\textwidth\epsfbox{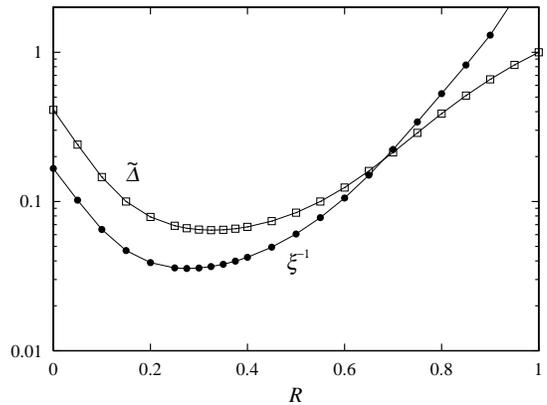}}
 \vspace*{1.0em}
 \caption{$R$-dependence of the inverse correlation length along the
 chains (solid circles) and the spin gap (open squares).  The
 statistical error of each data point is much smaller than the symbol
 sizes.}
 \label{fig:gap}
\end{figure}
 
The {\em non-existence} of phase transitions is also confirmed by the
$R$-dependence of the spin gap, $\tilde{\Delta}$, and the inverse
correlation length, $\xi^{-1}$ (Fig.~\ref{fig:gap}).  These results
convincingly support the conjecture made in the previous analytic
studies.~\cite{Senechal1995,AllenS2000} It should be recalled that by
using the method we explained in detail in Sec.~\ref{subsec:limit}, the
convergence to the thermodynamic and the zero-temperature limits of the
data has been checked for all the value of $R$ we simulated.  Therefore,
the data can be identified with those at $L=\infty$ and $T=0$ besides
the statistical errors, which are much smaller than the symbol sizes in
Figs.~\ref{fig:ssus} and \ref{fig:gap}.

Although there are no singularities between $R=0$ and 1, long-range
antiferromagnetic fluctuations are greatly enhanced in the intermediate
region of $R$.  This is consistent with the temperature dependence of
the uniform susceptibility presented in Sec.~\ref{subsec:usus}.
Especially, note that all the physical quantities we calculated have its
maximum (or minimum) at $R \simeq 0.3$ (see also
Table~\ref{tab:convergence}).  They could be compared with those of the
spin-2 antiferromagnetic Heisenberg chain, i.e., $\tilde{\chi}_{\rm s} =
1164.0(2)$, $\xi = 49.49(1)$, and $\tilde{\Delta} =
0.08917(4)$.~\cite{TodoK2001}

\section{Plaquette-Singlet Solid State and Hidden Order Parameter}
\label{sec:pss}

\subsection{Breakdown of AKLT picture}
\label{subsec:aklt}

As we have already mentioned in Sec.~\ref{sec:intro}, the ground state
of the spin-1 antiferromagnetic chain is understood quite well by means
of the VBS picture.~\cite{AffleckKLT1987} Actually, the AKLT state,
which is the exact ground state of the so-called AKLT
model,~\cite{AffleckKLT1987} shares many common properties, such as
spin-rotation symmetry, finite correlation length, etc., with the ground
state of the spin-1 Heisenberg chain.  These two states are believed to
belong to the same universality class with each other.
 
\begin{figure}[tb]
 \centerline{\epsfxsize=0.32\textwidth\epsfbox{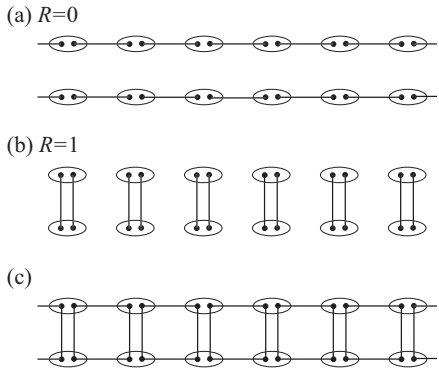}}
 \vspace*{1.5em}
 \caption{Schematic picture of plaquette-singlet solid state.  (a) AKLT
 state at $R=0$.  (b) Spin-1 dimer state at $R=1$ (c) Plaquette-singlet
 solid state for $0 < R < 1$.}
 \label{fig:pss}
\end{figure}

The AKLT state is essentially constructed as direct products of
spin-$\frac{1}{2}$ dimers sitting on each bond (Fig.~\ref{fig:pss}(a)).
On each site, two edge $S=\frac{1}{2}$ spins are symmetrized to form an
$S=1$ spin.  An important feature of this state is that a sum of $S^z$
in any interval $[i,j]$ can take only 1, 0, or -1.  This fact is
immediately followed by the existence of a topological hidden order; if
one removes spins at $S^z=0$ state, the sequence of $S^z$ of the
remaining spins has a perfect antiferromagnetic-like ordering (1, -1, 1,
-1, $\cdots$), although the original one has no `true' antiferromagnetic
long-range order at all.

The above topological order in the AKLT state can be detected
quantitatively by means of the {\em string order
parameter},\cite{denNijsR1989} which is defined by
\begin{equation}
 \label{eqn:ostr}
 \langle {\cal O}_2 \rangle = \lim_{|i-j| \rightarrow \infty}
 \langle {\cal O}_2(i,j) \rangle
\end{equation}
in terms of the string correlation operator:
\begin{equation}
 \label{eqn:cstr}
 {\cal O}_2 (i,j) = - S_{i}^z
  \exp \left[ i\pi \sum_{k=i+1}^{j-1} S_{k}^z \right] S_{j}^z \,,
\end{equation}
whose expected value is $\frac{4}{9} = 0.444\cdots$ for any $|i-j|>1$.
The string order parameter (\ref{eqn:ostr}) is also finite in the spin-1
Heisenberg chain.  Its value is estimated to be 0.3743(1) in the
thermodynamic limit ($R=0$ in Fig.~\ref{fig:hidden}).  In practice, the
string order parameter of a finite chain of length $L$ is calculated as
\begin{equation}
 \langle {\cal O}_2 \rangle_L = \frac{1}{L} \sum_{i} \langle 
  {\cal O}_2(i,i+L/2) \rangle \,.
\end{equation}
Although the string correlation function (\ref{eqn:cstr}) is a non-local
quantity, one can construct an improved estimator even for it.  For
details, see Ref.~\onlinecite{TodoKT2000}.

In the presence of the interchain coupling, however, the above AKLT
picture breaks down immediately.  As shown in Fig.~\ref{fig:hidden}, the
string order parameter, $\langle {\cal O}_2 \rangle_L$, defined on one
of the two chains, decreases quite rapidly as $R$ increases.
Furthermore, in contrast to the other quantities shown before, the value
of $\langle {\cal O}_2 \rangle_L$ exhibits strong system-size dependence
for $R>0$.  Actually, as shown in the inset of Fig.~\ref{fig:hidden},
one finds that these finite-size data scale {\em exponentially} quite
well as
\begin{equation}
 \langle {\cal O}_2 \rangle_L \simeq \tilde{f}(R \, \log^\alpha \! L)
\end{equation}
with $\alpha = 2.5$, where $\tilde{f}(x)$ is a scaling function, and it
vanishes for $x\rightarrow\infty$.  This strongly suggests that the
string order parameter (\ref{eqn:ostr}) is {\em essentially singular} at
$R=0$, and is vanished by infinitesimal interchain coupling, though we
have no account for the value of the exponent $\alpha$ at the moment.

\subsection{Plaquette-singlet solid state}
\label{subsec:pss}

As we have seen above, the AKLT picture (Fig.~\ref{fig:pss}~(a)) breaks
down immediately for $R>0$.  Actually, in the $R=1$ limit, the ground
state is represented schematically by a pattern of spin-$\frac{1}{2}$
dimers, in which two dimers are sitting on each rung
(Fig.~\ref{fig:pss}~(b)).  There is no overlap between these two VBS
states.

However, if we focus our attention on a plaquette containing four spins
at $(\alpha,x) = (1,i)$, $(1,i+1)$, $(2,i)$, and $(2,i+1)$, it is found
that these two states share a remarkable feature, i.e., the sum of $S^z$
of these four spins can take only 0, $\pm 1$, or $\pm 2$, and its
absolute value never exceeds 2 in both cases.  In other words, four
$S=\frac{1}{2}$ spins, each of which belongs to an $S=1$ spin at one of
four corners of a plaquette, form an $S=0$ state, though its fine
structure is quite different from each other.  This observation leads us
to a proposal of the following generalized state, which connects these
two VBS states smoothly.

First, we define a local state of plaquette consisting of four
$S=\frac{1}{2}$ spins, which is expressed explicitly as
\begin{eqnarray}
 \label{eqn:ps}
 |\psi(\theta)\rangle_i & = &
 a_\theta \Big[ \cos \theta 
 \left\{
    |\!\uparrow\rangle_{1,i} |\!\downarrow\rangle_{1,i+1}
  - |\!\downarrow\rangle_{1,i} |\!\uparrow\rangle_{1,i+1}
 \right\} \nonumber \\
 & & \quad \times
 \left\{
    |\!\uparrow\rangle_{2,i} |\!\downarrow\rangle_{2,i+1}
  - |\!\downarrow\rangle_{2,i} |\!\uparrow\rangle_{2,i+1}
 \right\} \nonumber \\
 & & \quad + \sin \theta 
 \left\{
    |\!\uparrow\rangle_{1,i} |\!\downarrow\rangle_{2,i}
  - |\!\downarrow\rangle_{1,i} |\!\uparrow\rangle_{2,i}
 \right\} \\
 & & \quad \times
 \left\{
    |\!\uparrow\rangle_{1,i+1} |\!\downarrow\rangle_{2,i+1}
  - |\!\downarrow\rangle_{1,i+1} |\!\uparrow\rangle_{2,i+1}
 \right\} \Big] \,, \nonumber
\end{eqnarray}
where $a_\theta$ is the overall normalization factor ($a_\theta^{-1} =
\sqrt{4+2\sin 2\theta}$).  This state is constructed as a superposition of
two $S=0$ states.  The first term of r.h.s. in Eq.~(\ref{eqn:ps}) is the
product of the dimers sitting on legs, and the second one does that of
the rung dimers.  The parameter $\theta$ ($0 \le \theta \le
\frac{\pi}{2}$) controls the proportion of these two constituents.  Note
that the wave function (\ref{eqn:ps}) is the exact ground state of the
following four-body spin-$\frac{1}{2}$ Hamiltonian:
\begin{eqnarray}
 \label{eqn:ph}
 {\cal H}_{\rm p} &=& \frac{J_{\rm p}}{4} (\sigma_{1,i} \cdot \sigma_{1,i+1} + \sigma_{2,i} \cdot \sigma_{2,i+1}) \nonumber \\
  && \,\,\,\, \mbox{} + \frac{K_{\rm p}}{4} (\sigma_{1,i} \cdot \sigma_{2,i} + \sigma_{1,i+1} \cdot \sigma_{2,i+1}) \,,
\end{eqnarray}
where $J_{\rm p} \ge 0$, $K_{\rm p} > 0$, and $\sigma_{\alpha,i}$
denotes the spin-$\frac{1}{2}$ Pauli operator.  In this case, the
parameter $\theta$ in Eq.~(\ref{eqn:ps}) can be expressed explicitly as
a function of $x_{\rm p} \equiv K_{\rm p}/J_{\rm p}$ as
\begin{equation}
 \theta = \arctan \left(x_{\rm p}-1+\sqrt{1-x_{\rm p}+x_{\rm p}^2}\right) \,,
\end{equation}
and the gap above the ground state never closes throughout while one
varies $x_{\rm p}$ from 0 ($\theta = 0$) to $\infty$ ($\theta =
\frac{\pi}{2}$), as we will see in Sec.~\ref{sec:summary}.

\begin{figure}[tb]
 \centerline{\epsfxsize=0.42\textwidth\epsfbox{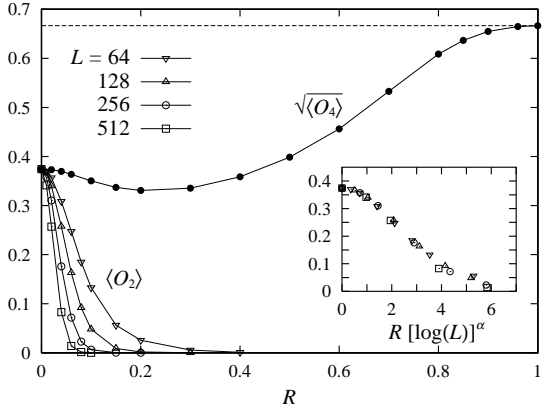}}
 \vspace*{1.0em}
 \caption{$R$-dependence of the hidden order parameters, $\langle O_2
 \rangle$ (open symbols) and $\protect\sqrt{\langle O_4 \rangle}$ (solid
 circles).  The statistical error of each data point is much smaller
 than the symbol sizes.  The horizontal dashed line denotes the exact
 value ($\frac{2}{3}$) of $\protect\sqrt{\langle O_4 \rangle}$ in the
 dimer limit ($R=1$).  A scaling plot for $\langle O_2 \rangle$ with
 exponent $\alpha=2.5$ is also shown in the inset.}
 \label{fig:hidden}
\end{figure}

By using the local singlet state (\ref{eqn:ps}), we construct a wave
function of the spin-1 ladder:
\begin{equation}
 \label{eqn:pss}
 |\Psi (\theta)\rangle_{\rm ps} = 
  \prod_{\alpha,i} {\cal P}_{\alpha,i} \prod_i |\psi(\theta)\rangle_i \,,
\end{equation}
where ${\cal P}_{\alpha,i}$ is a projection operator acting on two
$S=\frac{1}{2}$ spins at site $(\alpha,i)$.  The schematic picture of
this state is presented in Fig.~\ref{fig:pss}~(c).  We refer to it as
the {\em plaquette-singlet solid state}.  One sees that by varying the
parameter $\theta$ from 0 to $\frac{\pi}{2}$, it connects smoothly the
AKLT state at $R=0$ (Fig.~\ref{fig:pss}~(a)) and the spin-1 dimer state
at $R=1$ (Fig.~\ref{fig:pss}~(b)).  We expect the ground state of the
present system can be described well by the plaquette-singlet solid
state with $\theta$ tuned for each specific value of $R$.~\cite{pss}

Furthermore, it is possible to define a {\em four-body string
correlation operator}:
\begin{equation}
 \label{eqn:o4}
 {\cal O}_4 (i,j)  =  S_{1,i}^z S_{2,i}^z
  \exp \left[ i\pi \sum_{k=i+1}^{j-1}
		    \left( S_{1,k}^z + S_{2,k}^z \right) \right] 
  S_{1,j}^z S_{2,j}^z \,,
\end{equation}
which characterizes the plaquette-singlet solid state (\ref{eqn:pss}).
It is easy to prove that the expectation value of ${\cal O}_4(i,j)$
remains finite for any value of $\theta$ for the plaquette-singlet solid
state~(\ref{eqn:pss}).  Especially, $\langle {\cal O}_4 (i,j) \rangle =
\frac{4}{9}$ at $\theta = \frac{\pi}{2}$ ($R=1$).  It is interesting to
see that ${\cal O}_4(i,j)$ can be expressed as a product of the
conventional string correlations (\ref{eqn:cstr}) defined on each chain,
${\cal O}_2(i,j,\alpha)$:
\begin{equation}
 {\cal O}_4(i,j) = {\cal O}_2(i,j,1) \times {\cal O}_2(i,j,2) \,.
\end{equation} 
Therefore, in the decoupled-chain case ($R=0$) we have
\begin{equation}
 \label{eqn:relation}
 \langle {\cal O}_4(i,j) \rangle = \langle {\cal O}_2(i,j) \rangle^2 \,,
\end{equation} 
where we omit the index $\alpha$, since the expectation value of the
string correlation operator ${\cal O}_2(i,j;\alpha)$ does not depend on
$\alpha$.  For $R>0$, such a simple relation does not hold, and in
general they take different values with each other.

\begin{figure}[tb]
 \centerline{\epsfxsize=0.41\textwidth\epsfbox{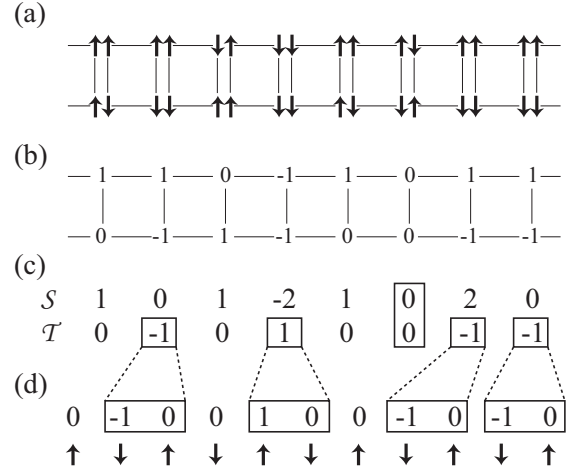}}
 \vspace*{1.5em}
 \caption{Demonstration of topological hidden order in the
 plaquette-singlet solid state. (a) Spin configuration on each
 plaquette.  Arrows denote the configuration ($\pm \frac{1}{2}$) of
 $S=\frac{1}{2}$ spins.  On each plaquette, the sum of configurations of
 four spins is restricted to be zero.  (b) Spin configuration of
 corresponding spin-1 ladder.  (c) Sequences $\{{\cal S}_i\}$ and
 $\{{\cal T}_i\}$ calculated from the configuration (b) as ${\cal S}_i =
 S^z_{1,i} + S^z_{2,i}$ and ${\cal T}_i = S^z_{1,i} S^z_{2,i}$,
 respectively.  (d) Sequence of ${\cal T}_i$'s obtained by being removed
 elements satisfying both of ${\cal T}_i = 0$ and $|{\cal S}_i| \ne 1$,
 and replaced 1 and -1 by $(1,0)$ and (-1,0), respectively, in which all
 positive (negative) ${\cal T}_i$'s sit on one (another) sublattice of
 the sequence.}
 \label{fig:topology}
\end{figure}

In Fig.~\ref{fig:hidden}, we show the $R$-dependence of our new hidden
order parameter,
\begin{equation}
 \langle {\cal O}_4 \rangle_L = \frac{1}{L} \sum_i \langle {\cal O}_4(i,i+L/2) \rangle \,,
\end{equation}
calculated for the present model.  We plot $\sqrt{\langle {\cal O}_4
\rangle_L}$ instead of $\langle {\cal O}_4 \rangle_L$ itself to
demonstrate the relation (\ref{eqn:relation}) at $R=0$.  In contrast to
the conventional string order parameter, $\langle {\cal O}_2 \rangle_L$,
which vanishes immediately for $R>0$, the new string order parameter,
$\langle {\cal O}_4 \rangle_L$, is found to be a smooth function of $R$,
and remains finite up to the dimer limit ($R=1$).

It should be emphasized that although the long-range spin fluctuations
are greatly enhanced round $R = 0.3$ as discussed in
Sec.~\ref{sec:results}, $\langle {\cal O}_4 \rangle_L$ remains
remarkably large even in this region.  This strongly supports that the
ground state of the spin-1 ladder is actually described quite well by
the plaquette-singlet solid state (\ref{eqn:pss}) in the whole range of
$R$.

\begin{figure}[tb]
 \centerline{\epsfxsize=0.42\textwidth\epsfbox{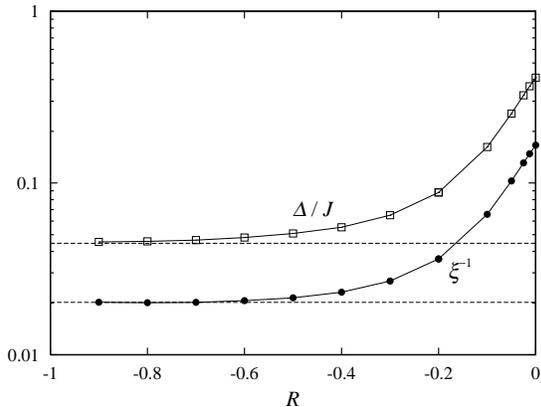}}
 \vspace*{1.0em}
 \caption{$R$-dependence of the spin gap (open squares) and the inverse
 correlation length along the chain (solid circles) in the ferromagnetic
 ($K<0$) case.  The dashed lines denote the respective values in the $R
 \rightarrow -1$ limit (Ref.~\protect\onlinecite{TodoK2001}).}
 \label{fig:gap-f}
\end{figure}

Before closing this section, it might be worth noting that the
plaquette-singlet solid state (Eq.~(\ref{eqn:pss}) and Fig.~\ref{fig:pss}~(c))
has the following {\em topological} antiferromagnetic long-range order,
though its definition is slightly complicated than its counterpart in
the AKLT state for the single spin-1 chain; first, for a given spin
configuration $\{S^z_{\alpha,j}\}$ ($\alpha=1,2$ and $i=1,2,3,\cdots$),
define two sequences, $({\cal S}_1, {\cal S}_2, {\cal S}_3, \cdots)$ and
$({\cal T}_1, {\cal T}_2, {\cal T}_3, \cdots)$, where ${\cal S}_i$'s and
${\cal T}_i$'s are calculated as ${\cal S}_i = S^z_{1,i} + S^z_{2,i}$
and ${\cal T}_i = S^z_{1,i} S^z_{2,i}$, respectively.  Next, choose a
nearest pair of non-vanishing ${\cal T}_i$'s, say ${\cal T}_j$ and
${\cal T}_k$.  If the number of ${\cal S}_\ell$'s satisfying $|{\cal
S}_\ell| = 1$ in the interval $j<\ell<k$ is even (odd), then ${\cal
T}_j$ and ${\cal T}_k$ has a same (opposite) sign.  In other words, in
the sequence of ${\cal T}_i$'s, if one removes elements satisfying both
of ${\cal T}_i = 0$ and $|{\cal S}_i| \ne 1$, and replaces 1 and -1 by
$(1,0)$ and (-1,0), respectively, then one finds the resulting sequence
$\{{\cal T}_i\}$ has an antiferromagnetic long-range order regarding
non-zero elements, i.e., all positive (negative) ${\cal T}_i$'s sit on
one (another) sublattice of the sequence.  A demonstration of this
topological order is presented in Fig.~\ref{fig:topology}.

\section{Ferromagnetic Interchain Coupling}
\label{sec:ferro}

So far, we consider only the case where the interchain coupling is
antiferromagnetic ($R>0$).  In this section, we consider in turn the
ferromagnetic case.  In the $R \rightarrow -1$ limit, two $S=1$ spins
connected by an infinitely-strong ferromagnetic rung bond form $S=2$.
Thus, two intrachain antiferromagnetic coupling of strength $J$ in
Eq.~(\ref{eqn:Hamiltonian}) are transformed into one bond of strength
$\tilde{J}=J/2$ connecting two $S=2$ spins.  The spin-2
antiferromagnetic chain has a finite spin gap, $\Delta/\tilde{J} =
0.08917(4)$, which is much smaller than that of the spin-1 chain, and
also a longer correlation length, $\xi = 49.49(1)$.~\cite{TodoK2001}

\begin{figure}[tb]
 \centerline{\epsfxsize=0.42\textwidth\epsfbox{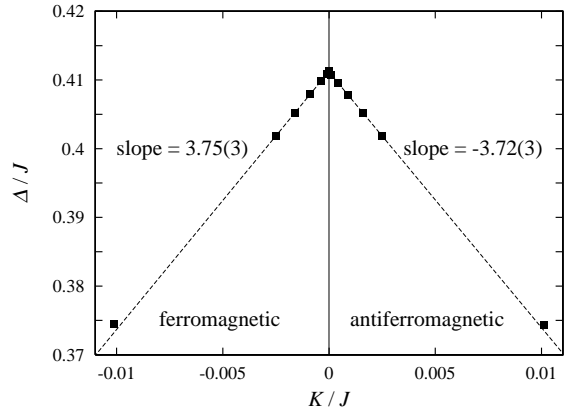}}
 \vspace*{1.0em}
 \caption{$K/J$-dependence of the spin gap, $\Delta/J$, in the
 antiferromagnetic ($K>0$) and ferromagnetic ($K<0$) cases.  The dotted
 straight lines are obtained by least-squares fitting for $0 \le K/J <
 0.025$ and $-0.025 < K/J \le 0$, respectively.}
 \label{fig:f-af}
\end{figure}
 
The spin-2 Haldane state is also explained by a VBS picture, in which
two dimers sitting on each bond.  Therefore, one would expect naturally
that the two independent spin-1 Haldane state at $R=0$ and the spin-2
one ($R \rightarrow -1$) is connected smoothly without any singularities
between them.  Indeed, as shown in Fig.~\ref{fig:gap-f}, the spin gap,
$\Delta/J$, as well as the correlation length, $\xi$, decreases
monotonically as $R$ does, and seems to converge smoothly to the values
of the spin-2 Haldane chain.  Note that the spin gap presented in
Fig.~\ref{fig:gap-f} is not the reduced gap, $\tilde{\Delta}$.  Since
$\tilde{\Delta}$ vanishes as $\sim 1/|K|$ for $R \ll 0$, we normalize
the gap as $\Delta/J$, which remains finite in the $R \rightarrow -1$
limit.  At $R=-0.8$ and $-0.9$, we need $L=512$ and
$\tilde{T}=0.0009766$ to obtain the zero-temperature values in the
thermodynamic limit.

Although the decoupled-chain point ($K=0$) is not a critical point, the
spin gap is still {\em non-analytic} at this point.  In
Fig.~\ref{fig:f-af}, we show the $K/J$-dependence of $\Delta/J$ near
$K=0$ ($|K/J| \le 0.01$) both in the antiferromagnetic and ferromagnetic
regimes.  One sees a clear cusp just at $K=0$.  In both regimes,
$\Delta/J$ decreases linearly as $|K/J|$ increases.  Furthermore, we
find the absolute value of its slope is the same within the error on
both sides (3.72(3) for $K>0$ and 3.75(3) for $K<0$).  We should remark
this is completely consistent with the conjecture by the previous
bosonization study.~\cite{AllenS2000}

\section{Summary and Discussions}
\label{sec:summary}

In this paper, we investigated precisely the ground-state properties of
the spin-1 antiferromagnetic Heisenberg two-leg ladder by means of the
quantum Monte Carlo simulation.  We found that the system is gapful
regardless of the strength of interchain coupling, that is, the Haldane
state in the decoupled chains and the spin-1 dimer state are connected
smoothly and there is no quantum phase transition between them.  We
conclude that the behavior of the spin gap as a function of $R$ we
observed, including the cusp at $R=0$, is completely consistent with the
recent analytic studies based on the mapping to the nonlinear $\sigma$
model~\cite{Senechal1995} and the bosonization
technique.~\cite{AllenS2000}

Although there is no quantum phase transition in the present system, the
spin gap is greatly suppressed, and at the same time the long-range
antiferromagnetic fluctuations are enhanced in the intermediate region
($R \simeq 0.3$), e.g., $\Delta/(J+K) = 0.06476(4)$ and $(J+K) \chi_{\rm
s} = 490.1(2)$ at $R=0.3$.  This explains the reason why the spin-1
ladder with $R \simeq 0.3$ exhibits an antiferromagnetic long-range
order, if one introduces quite small ($\simeq 0.004J$) antiferromagnetic
coupling between ladders.~\cite{Matsumoto2000} It should be noted that
the critical inter-ladder coupling is smaller by an order of magnitude
than the spin gap.  This situation is the same as the critical
interchain coupling, $J'_{\rm c}$, of the two-dimensional array of
spin-1 chains, $J'_{\rm c}/J =0.043648(8)$,~\cite{Matsumoto2000} while
$\Delta/J = 0.41050(2)$.~\cite{WhiteH1993} The precise ground-state
phase diagram of these two-dimensional spin-1 Heisenberg
antiferromagnets has been reported in detail in
Ref.~\onlinecite{Matsumoto2000}.

The AKLT picture, which works quite well for the single spin-1 chain,
was found to break down immediately in the presence of interchain
coupling.  We proposed the plaquette-singlet solid state to describe the
ground state qualitatively for the whole range of $R$.  The
plaquette-singlet solid state has a topological hidden order.  The
hidden order parameter $\langle {\cal O}_4 \rangle$, which characterizes
the plaquette-singlet solid state, was shown to be finite up to the
dimer limit ($R=1$).  This strongly supports that the ground state of
the spin-1 ladder is described quite well by the present
plaquette-singlet solid picture.

It is quite interesting to see that the local Hamiltonian of
spin-$\frac{1}{2}$ plaquette, Eq.~(\ref{eqn:ph}), itself can reproduce
qualitatively correct behavior of the spin gap observed in the present
spin-1 ladder.  The Hamiltonian (\ref{eqn:ph}) has a singlet ground
state regardless of the value of $x_{\rm p} \equiv K_{\rm p}/J_{\rm p}$
($-\infty \le x_{\rm p} \le \infty$).  The $x_{\rm p}$-dependence of
the spin gap can be written explicitly as follows:
\begin{equation}
 \frac{\Delta}{J_{\rm p}} = \left\{
  \begin{array}{ll}
   \sqrt{1-x_{\rm p}+x_{\rm p}^2} & \,\,\,\, \mbox{for $x_{\rm p} \ge 0$} \\
   x_{\rm p}+\sqrt{1-x_{\rm p}+x_{\rm p}^2} & \,\,\,\, \mbox{for $x_{\rm p} \le 0.$} \\
  \end{array}
  \right.
\end{equation}
One can find easily the following features: (i) There always exists a
finite gap for $-\infty \le x_{\rm p} \le \infty$. (ii) For $x_{\rm
p}>0$ ($K_{\rm p}>0$), $\Delta/(J_{\rm p}+K_{\rm p})$ has a minimum at a
finite value of $x_{\rm p}$. (iii) On the other hand, for $x_{\rm p}<0$
($K_{\rm p}<0$), $\Delta/J_{\rm p}$ is a monotonically-decreasing
function of $|x_{\rm p}|$, and converges to a finite value at $x_{\rm
p}=-\infty$.  (iv) Finally, at $x_{\rm p}=0$, $\Delta/J_{\rm p}$ has a
cusp, and it decreases linearly with the same slope on both sides.  The
last behavior is simply due to the crossing of the second- and
third-lowest eigenvalues.  All of these features are qualitatively the
same as the present spin-1 ladder.

The plaquette-singlet solid picture proposed in the present paper might
be generalized to wider classes of the spin-1 Heisenberg ladder.
Indeed, for the ferromagnetic interchain coupling case discussed in
Sec.~\ref{sec:ferro}, our hidden order parameter $\langle O_4 \rangle$
was found to remain at a finite value ($\simeq 0.089$) in the limit $K
\rightarrow -\infty$.  Furthermore, the present plaquette-singlet solid
picture might be applied even in the presence of bond alternation
(forced dimerization) in the intrachain coupling, i.e., $J_i = (1 +
(-1)^i \delta) J$, where $\delta$ denotes the strength of dimerization.
We have observed that there is no singularity for the whole range of
$\delta$ ($0 \le \delta \le 1$), and $\langle O_4 \rangle$ remains
finite up to the decoupled-plaquette limit ($\delta = 1$).  Note that as
for the decoupled dimerized chain ($K=0$), the $\delta$-dependence of
$\langle O_4 \rangle$ is qualitatively different from that for $K>0$.
There exists a critical point at $\delta_{\rm c} =
0.26001(4)$,~\cite{NakamuraVT2001} and the conventional string order
parameter $\langle O_2 \rangle$ vanishes for $\delta \ge \delta_{\rm
c}$.  Since the relation (\ref{eqn:relation}) holds for $K=0$, the new
hidden order parameter $\langle O_4 \rangle$ also vanishes for $\delta
\ge \delta_{\rm c}$.  The existence of non-zero hidden order parameter
$\langle O_4 \rangle$ in the cases with ferromagnetic couplings and with
alternating bonds implies their ground states are also well described by
products of local singlet states similar to the present
plaquette-singlet solid state, though the structure of local singlet
state should be quite different from the present one.
  
Very recently a new nitroxide material, abbreviated as BIP-TENO, has
been synthesized, and its magnetic properties have been investigated
precisely.~\cite{KatohHIG2000,GotoBHKI2001} The tetraradical BIP-TENO
molecule consists of two pairs of ferromagnetically-coupled
$S=\frac{1}{2}$ spins, and relatively weak antiferromagnetic coupling
exists between the pairs.  The crystalline state of BIP-TENO is thus
expected to be described effectively by a spin-1 antiferromagnetic
ladder of present interest.  From the high-temperature behavior of the
uniform susceptibility, Katoh {\em et al.} estimated the strength of
effective couplings and concluded that $J \simeq 50{\rm K}$ and $K
\simeq 42{\rm K}$, i.e., $R \simeq 0.46$.\cite{KatohHIG2000} A clear
spin-gapped behavior of the uniform susceptibility observed in the
experiment is consistent with the present results.  The excitation gap
$\Delta$ is also estimated to be 15.6K from the magnetization
curve.~\cite{GotoBHKI2001} Unfortunately, its magnitude is more than
twice as large as the present result ($\Delta/(J+K) = 0.07392(6)$ at
$R=0.45$.  See Fig.~\ref{fig:gap}).

Interestingly, the shoulder in the temperature dependence of the uniform
susceptibility (Fig.~\ref{fig:usus}) has also been observed even in the
real material.~\cite{KatohHIG2000} This anomaly in the excitation
spectrum might be related with the $\frac{1}{4}$-plateau observed in the
magnetization process of BIP-TENO.~\cite{GotoBHKI2001} However, it is
beyond the scope of the present study, and remains as a future problem.

\section*{Acknowledgement}

One of the present author (S.T.) thanks K. Totsuka for reminding him the
previous works on the present subject.  The computation in the present
work has been performed mainly on the 384~CPU massively-parallel
supercomputer, SGI 2800, at the Supercomputer Center, Institute for
Solid State Physics, University of Tokyo.  The program used in the
present simulation was based on the library `Looper version 2' developed
by S.T. and K. Kato and also on the `PARAPACK version 2' by S.T.  The
present work was supported by the ``Research for the Future Program''
(JSPS-RFTF97P01103) of the Japan Society for the Promotion of Science.
S.T.'s work was partly supported by the Swiss National Science
Foundation.


\begin{thebibliography}{99}
\bibitem{z} Here, we assume the Lorentz invariance, i.e., the dynamical
	 exponent $z$ is unity.
\end{thebibliography}
\end{document}